\documentclass{aa}

\usepackage{amsmath}
\usepackage{siunitx}
\usepackage{xfrac}
\usepackage{natbib}
\usepackage{xcolor}
\usepackage{graphicx}
\usepackage{booktabs}
\usepackage{csquotes}
\usepackage{subcaption}
\usepackage{multirow,array}
\makeatletter
  \renewcommand*\aa@pageof{, page \thepage{} of \pageref*{LastPage}}
\makeatother
\usepackage{hyperref}            

\hypersetup{pdfpagemode = {UseNone},
            pdftitle = {Circumbinary Planets},
            pdfcreator = {\LaTeX},
            pdfproducer = {pdfeTeX-0.\the\pdftexversion\pdftexrevision},
            pdfauthor = {Anna Penzlin, Wilhelm Kley, Richard Nelson},
            pdfsubject = {},
            pdfview = {FitH},
            pdfstartview = {FitH},
            colorlinks = {true},
            linkcolor = [rgb]{0,0.35,0.7},
            citecolor = [rgb]{0,0.35,0.7},
            filecolor = [rgb]{0.61,0,0},
            urlcolor = [rgb]{0.61,0,0},
           }

\usepackage[normalem]{ulem}
           
\usepackage{bm}

\colorlet{darkgreen}{green!60!black}

\colorlet{darkpink}{magenta!90!black}

\newcommand{\beq}{\begin{equation}}
\newcommand{\eeq}{\end{equation}}

\begin{document}

\title{Parking planets in circumbinary discs}

\author{Anna~B.~T. Penzlin\inst{\ref{inst1}},
	Wilhelm Kley\inst{\ref{inst1}},
    Richard~P. Nelson\inst{\ref{inst2}}}

\institute{
Institut f\"ur Astronomie und Astrophysik, Universität T\"ubingen,
Auf der Morgenstelle 10, D-72076, Germany \label{inst1}
\and Astronomy Unit, School of Physics and Astronomy, Queen Mary University of London, London E1 4NS, UK \label{inst2}}

\date{}

\abstract
{
The Kepler space mission discovered about a dozen planets orbiting around
binary stars systems. Most of these circumbinary planets lie near
their instability boundaries at about 3 to 5 binary separations. Past attempts to match these final
locations through an inward migration process were only successful for the Kepler-16 system.
Here, we study 10 circumbinary systems and try to match the final parking locations
and orbital parameters of the planets with a disc driven migration scenario.

We performed 2D locally isothermal hydrodynamical simulations of circumbinary discs with embedded planets and followed their migration evolution
using different values for the disc viscosity and aspect ratio.
We found that for the six systems with intermediate binary eccentricities ($0.1 \le \mathrm{e_{bin}}\le 0.21$)
the final planetary orbits matched the observations closely for a single set of disc parameters, specifically
a disc viscosity of $\alpha = 10^{-4}$, and an aspect ratio of $H/r \sim 0.04$.
For these systems the planet masses were large enough to open at least a partial gap in their discs as they approach the binary, forcing the discs to become circularized and allowing for further migration towards the binary, leading to good agreement with the observed planetary orbital parameters. 

For systems with very small or large binary eccentricities the match was not as good because
the very eccentric discs and large inner cavities in these cases prevented close-in planet migration.
In test simulations with higher than observed planet masses better agreement could be found for those systems.

The good agreement for 6 out of the 10 modelled systems, where the relative difference between observed and simulated final planet orbit is $\leq 10\%$, strongly supports the idea that planet migration in the disc brought the planets to their present locations.
}


\maketitle

\section{Introduction}\label{sec:intro}
The first circumbinary planet (CBP) discovered by the Kepler space mission was Kepler-16b \citep{2011Kepler16}. Since then about a dozen more circumbinary p-type planets have been discovered (see overview in \citet{2019MNRAS.488.3482M} and Table \ref{tab:Kep-data}). 
In 2012, multiple CBPs around Kepler-47 were detected \citep{2012Kepler47,2019Kepler47}.  
Later, the Planet Hunters project discovered a planet in a quadruple star system PH1b, also named Kepler-64b \citep{2013Kepler64}.
The most recent discovery is TOI-1338b by the TESS mission \citep{2020TOI-1338}. 

All of the above systems, and most of the discovered innermost CBPs, share a set of properties in their orbits. Firstly, they are in good alignment with the binary's orbital plane, hinting at formation in a protoplanetary disc that was aligned with the binary. 
Secondly, the planetary orbits (the semi-major axes) are all near 3.5 binary separations, which is close to the instability limit \citep{1986Dvorak}.
Additionally, the eccentricities of the planetary orbits are small.
All of this points to a common formation mechanism for most observed CBPs in aligned, coplanar discs.

Concerning the general stability of circumbinary discs it has been suggested that, in addition to the coplanar situation,
the perpendicular case is also a stable binary-disc configuration \citep{2017ApJ...835L..28M,2018MNRAS.473.3733L,2019Discalignment}. 
However, no planets have been observed in perpendicular orbits yet, due to observational constraints. 

In-situ formation of planets at these close orbits around binaries are unlikely. 
Tidal forces close to the binary cause perturbations in the inner disc that inhibit planetesimal and dust accretion \citep{2012Paardekooper,2015Silsbee,2012Meschiari}, and turbulence induced by hydrodynamical parametric instabilities can dramatically reduce pebble accretion efficiencies \citep{2020Pierens}.
%
A more likely formation scenario is planet formation in the outer disc and subsequent disk-driven migration to the observed close orbits \citep{2008Pierens,2015Bromley}. However, this does not in itself explain the particular stopping positions of most CBPs. 

\begin{table*}[t]
\centering
\begin{tabular}{|c|c|c|c|c|c|c|c|c|c|c|}
\hline 
Kepler & 38 & 35 & 453 & 64 & 34 & 47 & 16 & 413 & 1661 &TOI-1338\\ 
\hline 
$\mathrm{M_{bin}} [\mathrm{M_{sun}}]$&1.20&1.70&1.14&1.77&2.07&1.32&0.89&1.36&1.10&1.34\\
\hline 
$\mathrm{q_{bin}}$& 0.26& 0.91& 0.21& 0.28& 0.97& 0.35& 0.29& 0.66& 0.31& 0.28\\ 
\hline 
$\mathrm{e_{bin}}$& 0.10& 0.14& 0.05& 0.21& 0.52& 0.02& 0.16& 0.04& 0.11& 0.16\\ 
\hline 
$\mathrm{m_{p}} [\mathrm{m_{jup}}]$& 0.38& 0.12& 0.05& 0.1& 0.22& 0.05& 0.33& 0.21& 0.05& 0.10\\ 
\hline 
$\mathrm{a_{p}} \,[\mathrm{a_{bin}}]$& 3.16& 3.43& 4.26& 3.64& 4.76& 3.53& 3.14& 3.55& 3.39& 3.49\\ 
\hline 
$\mathrm{e_{p}}$& 0.03& 0.04& 0.04& 0.05& 0.18& 0.03& 0.006& 0.12& 0.06& 0.09\\ 
\hline 
$\mathrm{m_{p}}/\mathrm{M_{bin}}$ [$\times 10^{-5}$]  & 30 & 7.1 & 4.3 & 5.4 & 10 & 3.7 & 35 & 15 & 4.6 & 6.8 \\ 
\hline 
\end{tabular} 
\caption{Table of observed circumbinary parameters. $\mathrm{M_{bin}}$, $\mathrm{q_{bin}}$ and $\mathrm{e_{bin}}$ are the total mass, the mass ratio ($M_2/M_1$) and the eccentricity of the binary star, $\mathrm{m_{p}}$, $\mathrm{a_{p}}$, $\mathrm{e_{p}}$ are the mass, semi-major axis and eccentricity of the planet on its orbit. (See reference \cite{2012Kepler38}, \cite{2012Kepler34-35}, \cite{2015Kepler453}, \cite{2013Kepler64}, \cite{2019Kepler47}, \cite{2011Kepler16}, \cite{2014Kepler413}, \cite{2020TOI-1338}, \cite{2020Socia})
}
\label{tab:Kep-data}
\end{table*}

Several studies were already dedicated to investigate planet migration in circumbinary discs \citep{2008Pierens,2013Pierens,2014Kley,2015Kley,2019Kley}. Due to gravitational torques the binary clears out an inner cavity and hence planet migration is stopped at the inner edge of the disk at a location close to the binary, but not yet matching the observations in most cases studied.
Numerical simulations showed that the inner disc's shape and hence the final planet orbital parameters depend 
on disc parameters such as viscosity and scale height, and the mass ratio and eccentricity of the binary
\citep{2017Mutter,2018Thun}. 

Two recent studies by \citet{2017Mutter} and \citet{2019Kley} were able to reproduce the observed CBP orbits for Kepler-16b and Kepler-38b. In this work we plan to find a common formation scenario for most observed CBPs using a sample of 10 planets, as described in Table \ref{tab:Kep-data}.

To limit the parameter space, we will consider discs, in agreement with the low $\alpha$-viscosity that was observed in the D-sharp survey \citep{2018Dsharp6} and scale heights comparable to the results of \cite{2019Kley}. There, we showed also that the geometry of the disc,
specifically the eccentricity of the inner cavity, can be influenced by an embedded planet. Gap opening planets will circularize the disc
while lighter planets or discs with higher viscosity will have eccentric holes. We will explore the effect of a wider disc parameter range on the disc architecture in a future study.

To study the parking of planets we performed numerical hydrodynamical simulations of circumbinary discs using 3 different parameter sets for the disc physics in order to find a likely combination of viscosity and scale height that results in disc shapes that allow for suitable stopping
locations of the planets. Hence, this study extends the work of \citet{2013Pierens} where they looked at a subset of 3 systems.
The binary and planet parameters used in our simulations are those given by the observations, as presented in Table\,\ref{tab:Kep-data}.

In Section \ref{sec:methods} the adopted model assumptions and initial conditions are defined. Section \ref{sec:results} shows the results of the simulations, where we discuss the influence of viscosity on the planet migration in general and show the effects of different sets of disc parameters on the final orbits of the observed Kepler planets. We will discuss the difficulties of reaching the final orbit and the limitations of our model in Section~\ref{sec:dicussion} and summarise the findings.

\section{Simulations}\label{sec:methods}

We used the Pluto Code \citep{2007Mignone} in the GPU version by Daniel Thun \citep{2017Thun} to simulate 2D hydrodynamic disc models. We chose a cylindrical grid stretching from 1 to $40\,\mathrm{a_{bin}}$ with 684 logarithmically spaced radial grid cells and 584 azimuthal cells. The general design of the simulation follows the setup of \cite{2018Thun}. 

We modelled nine of the known Kepler binary planet systems, and TOI-1338, for which we used the observed binary and planet properties, as displayed in Table\,\ref{tab:Kep-data}. 

For the surface density profile we use $\Sigma \propto R^{-3/2}$ with a disc mass of $\mathrm{M_{d}}=0.01\,\mathrm{M_{bin}}$.
To speed up the equilibration process for the disc we initialized the surface density with an inner cavity reaching to about $2.5\, \mathrm{a_{bin}}$. Inside of this region the disc will be cleared due to the binary's action. 

All discs are evolved initially for $>25\,000$ binary orbits, $\mathrm{T_{bin}}$, without a planet being present, depending on the specific system setup.A typical time scale for a locally isothermal protoplanetary disc around a binary to reach a convergent quasi-equilibrium state is a few $10\,000\, T_\mathrm{bin}$. Hence, we used in this work locally isothermal models as they converge faster than viscously heated radiative disc which can need up to $200\,000\, T_\mathrm{bin}$, but eventually reach comparable final states with a pressure profil consistent with an unflared inner disc that has a constant aspect ratio between 0.05 and 0.03 depending on the system, see \citet{2019Kley} for details.
During this initialization phase the binary is not influenced by the disc.
As the final planetary orbits will depend on disc properties, we simulated discs for the observed systems in Table~\,\ref{tab:Kep-data} for different
values of $h$ and $\alpha$. The viscously heated, radiative models in \citet{2019Kley} resulted in an unflared inner disc. Therefore we chose here a constant aspect ratio, $h = H/R$.
Based on our previous experience \citep{2018Thun,2019Kley}, we chose the following three combinations: $h=0.04, \alpha=10^{-3};  h=0.04, \alpha=10^{-4}$; and $h=0.03, \alpha=10^{-4}$, only for Kepler-38 we used additionally $h=0.05$.

After the disc reached a convergent behaviour we added a planet with the observed mass (see Table\,\ref{tab:Kep-data}) in the outer disc at a few $\mathrm{a_{bin}}$ beyond the maximum of the surface density in the disc. 
In cases where only a upper planet-mass limit was observed like Kepler-38, we used the observed upper limit. This might overestimate the actual mass of a planet, but is still in aggrement with observations. In the whole sample the planet masses range from $0.05\,M_\mathrm{jup}$ to $0.38\,M_\mathrm{jup}$
The disc exerts gravitational forces on the planet which will start migrating through the disc.
The disc forces acting on the central binary star are also switched on at this time, but during the time span of the simulations the binary elements do not
change substantially. 
  
\begin{figure*}[htb]
    \centering
    \includegraphics[width=0.9\textwidth]{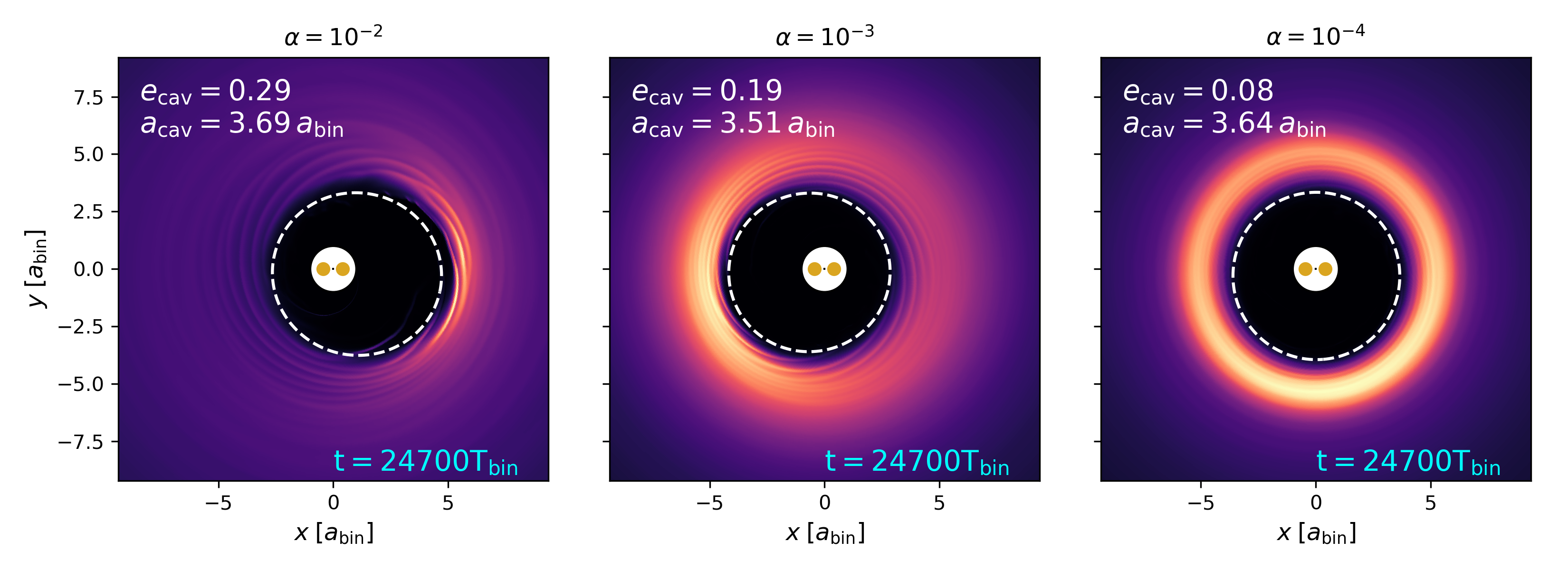}
    \caption{Surface density for discs around a Kepler-35 like binary for different $\alpha$-values and an aspect ratio of 0.04.
   The dashed white line denotes a fitted ellipse \citep{2018Thun} which marks the edge of the inner cavity with the parameters displayed in the top left corner. 
   }
    \label{fig:disc_alpha}
\end{figure*}

\begin{figure*}[htb]
    \centering
    \includegraphics[width=0.9\textwidth]{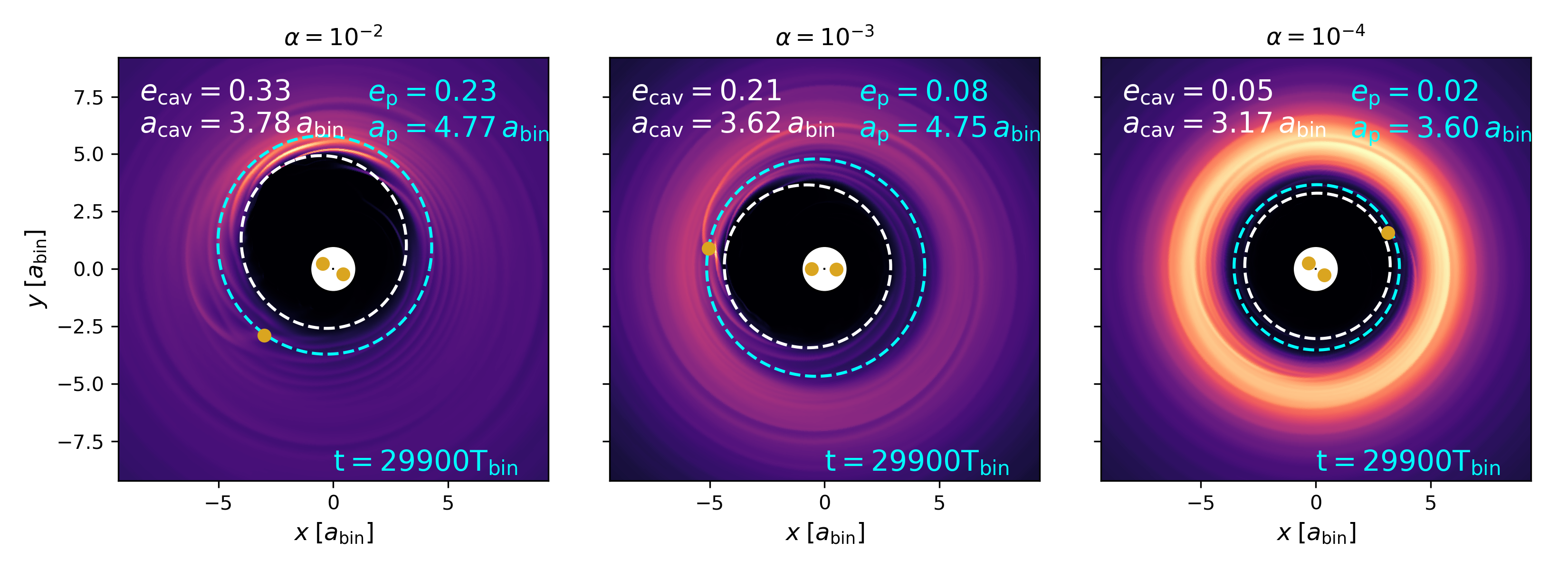}
    \caption{Discs for a Kepler-35 like system with different $\alpha$-values with a planet embedded for 29\,900 binary orbits. The dashed white line marks the edge of the inner cavity with the parameters displayed in the top right corner. The light blue line marks the orbits of the planet with the parameter displayed in the top left corner.
   }
    \label{fig:planet_alpha}
\end{figure*}

\section{Results}\label{sec:results}

In this section we present the results of our numerical studies of the systems quoted in
Tab.~\ref{tab:Kep-data}. We will first present results for different viscosities
on the specific system Kepler-35, and then concentrate on the complete set of systems.
An overview of the disc shapes that resulted in planetary evolutions that matched best with the observations
is given below in Fig.~\ref{fig:discs}.

\begin{figure*}[htb]
    \centering
    \includegraphics[width=0.9\textwidth]{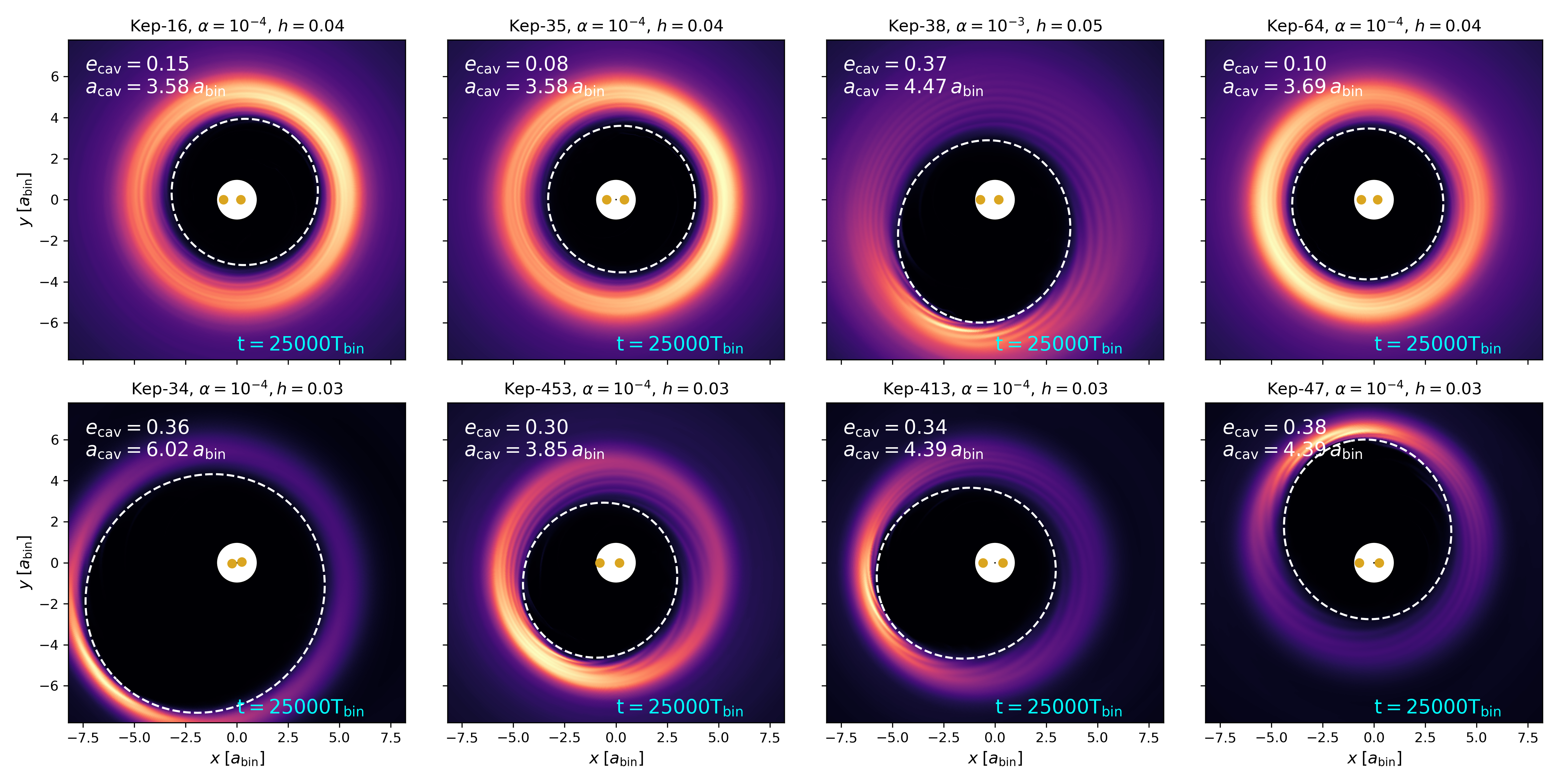}
    \caption{The surface density of 8 modelled Kepler circumbinary systems, listed in Table \ref{tab:Kep-data}, before the planet was embedded. The chosen $h$ and $\alpha$ correspond to those models that eventually could reproduce the observed final planetary orbits best. The white dashed ellipse marks the inner cavity edge with the parameters shown in the top left.}
    \label{fig:discs}
\end{figure*}

\begin{figure}[htb]
    \centering
    \includegraphics[width=0.5\textwidth]{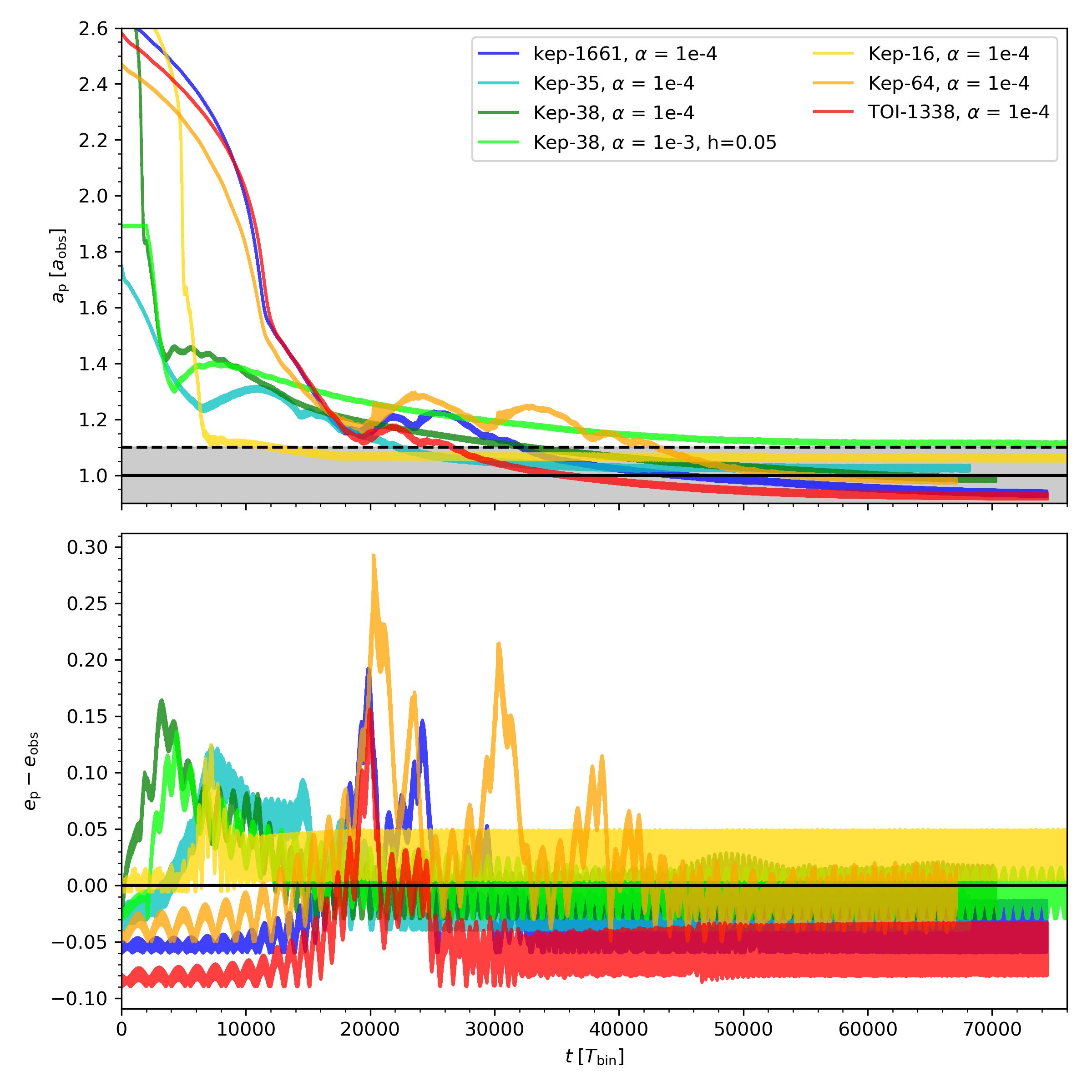}
    \caption{Migration history for all planets that get close to the observed orbit. Semi-major axis and eccentricity are scaled with the observed $\mathrm{a_{obs}}$ and $\mathrm{e_{obs}}$, see Tab.\,\ref{tab:Kep-data}. The grey area marks the success criterion,
    where the final parking position of the planet deviates by no more than 10\% from the observed value,
    i.e. $\Delta a_p/a_{p,obs} \leq 10\%$.
   For the discs we chose $\alpha=10^{-4}$ and $h=0.04$, plus one additional model for Kepler-38 using the stated values.
   }
    \label{fig:close_planets}
\end{figure}
\subsection{The influence of viscosity}

To analyse in more detail the impact of viscosity on the inner cavity we present results on the Kepler-35 system in more detail,
first the disc structure without a planet and then the impact of the planet.
The binary system Kepler-35 has a mass ratio close to unity, $\mathrm{q_{bin}}=0.91$, and a relatively small eccentricity of $\mathrm{e_{bin}} = 0.14$
which puts the system on the lower branch of the bifurcation diagram in the precession period versus cavity size graph \citep{2018Thun}.
In Figure\,\ref{fig:disc_alpha} equilibrium discs with different $\alpha$-values are displayed for a given scale height $h=0.04$
to show how much the viscosity influences the disc and cavity structure. The cavity parameters are calculated as described in \citet{2017Thun}, meaning they correspond to the parameters of an ellipse fitted to the location where the azimuthally averaged density is 10\% of the maximum density. The cavity and planet orbital parameters are calculated using Jacobi coordinates for the planet-binary system. 
The disc eccentricity decreases with decreasing viscosity, while the sharpness of the density maximum increases. As all observed planets reside on orbits with relatively low eccentricity,
low eccentric cavities will be favourable for their formation but the presence of the planet will also change the geometry
of the disc, as discussed in \cite{2017Mutter} and \cite{2019Kley}. Planets with orbits near the inner edge of the disc can circularize the cavity and disc. But high disc eccentricities, as they occur for binaries on circular and high eccentric orbits \citep{2018Thun}, can hinder a planet from reaching this orbit. Therefore, lower viscosities leading to less eccentric discs are favourable to explain the planet migration into the observed orbits.

A planet with the observed mass of Kepler-35 ($0.12\, \mathrm{m_{jup}}$) is embedded in the disc at an initial distance of $6\,\mathrm{a_{bin}}$ on a circular orbit and allowed to migrate in the disc.
Initially, the embedded planet follows the eccentric gas flow in the disc which makes its orbit eccentric. Planets that do not open a gap in the disc remain embedded and are pushed back by the overdensity at the cavity (see Fig.\,\ref{fig:disc_alpha}) to a parking position on an eccentric orbit within the disc. This case is shown in the left panel of Fig.~\ref{fig:planet_alpha} where the high disc viscosity
does not allow for a planetary gap to be cleared.

However, if planet and disc parameters are such that the planet can open a gap (high planet mass and low disc viscosity and temperature as discussed in \cite{2006Crida}) the disc will be
circularized and it can pass through the cavity overdensity to a more circular orbit inside of the cavity, as shown in \citet{2017Mutter, 2019Kley}.
Kepler-35b and all other observed planets have masses much lower than a Jupiter mass which puts constraints on the 
disc parameters that are linked to gap opening, such as viscosity and aspect ratio.

In Figure \ref{fig:planet_alpha} the final position and orbit of the planet in these discs is displayed. The gradual deepening of the planetary gap is evident between $\alpha=10^{-2}$ to $\alpha=10^{-3}$. Even though the gap in the middle panel is only partially opened
it already leads to a reduction in orbital eccentricity of the cavity and planet. However, in both cases the planet is still too well embedded in the disc and influenced by its dynamics such that the orbit of the planet  and the cavity are fully aligned and precess at the same rate, as also observed in \cite{2019Kley,2019Penzlin}. 
For an even lower viscosity of $\alpha=10^{-4}$ (right panel) the planet opens a deeper gap that erases directly the disc driven planet dynamics and allows the planet to reach a stable orbit inside the cavity. 
Therefore, a low disc viscosity 
tends to have a positive effect on disc structure and planet migration,
because this favours final planet orbits with low eccentricity, as observed in many systems.
Colder and less viscous discs have on average a less eccentric cavity. 
This aids the passage of the planet through the inner cavity. 

\subsection{Kepler systems}
After having demonstrated the main impact of the viscosity on the disc structure and outcome of the planet orbit
for Kepler-35, we present now the results for the other systems.
Instead of presenting results of the general impact of disc and binary parameters on the cavity dynamics,
we focus here on just that parameter set for which the simulations most successfully reproduced the observed planetary orbits, and which is reasonable for a protoplanetary disc. We will defer analysis of a broader parameter range to future work.

In Fig.\,\ref{fig:discs} we display the disc structure for models that most successfully could reproduce the planetary orbits.
The discs are shown at a time {\itshape before} the planet was embedded.
As suggested by the results for Kepler-35, discs with low viscosity and also small aspect ratios are favourable in producing
the observed orbits, as this combination of disc parameters will ensure that the planet opens a gap
at its orbit which results in a more circular disc cavity that allows the planet to migrate closer to the binary.
In the top row we have shown examples where the final planet orbits matched the observed locations well, while in the lower
panel the agreement was not as good. Indeed, the
main difference between the "successful" top row and the "unsuccessful" bottom row is the cavity's eccentricity. While the discs in the top row are typically more circular even without any planet, the bottom row discs are noticeably eccentric even for low $\alpha$-values, due to the small (Kepler-47, Kepler-413, and Kepler-453) or high (Kepler-34) binary eccentricities.

\begin{figure*}[htb]
    \centering
    \includegraphics[width=0.9\textwidth]{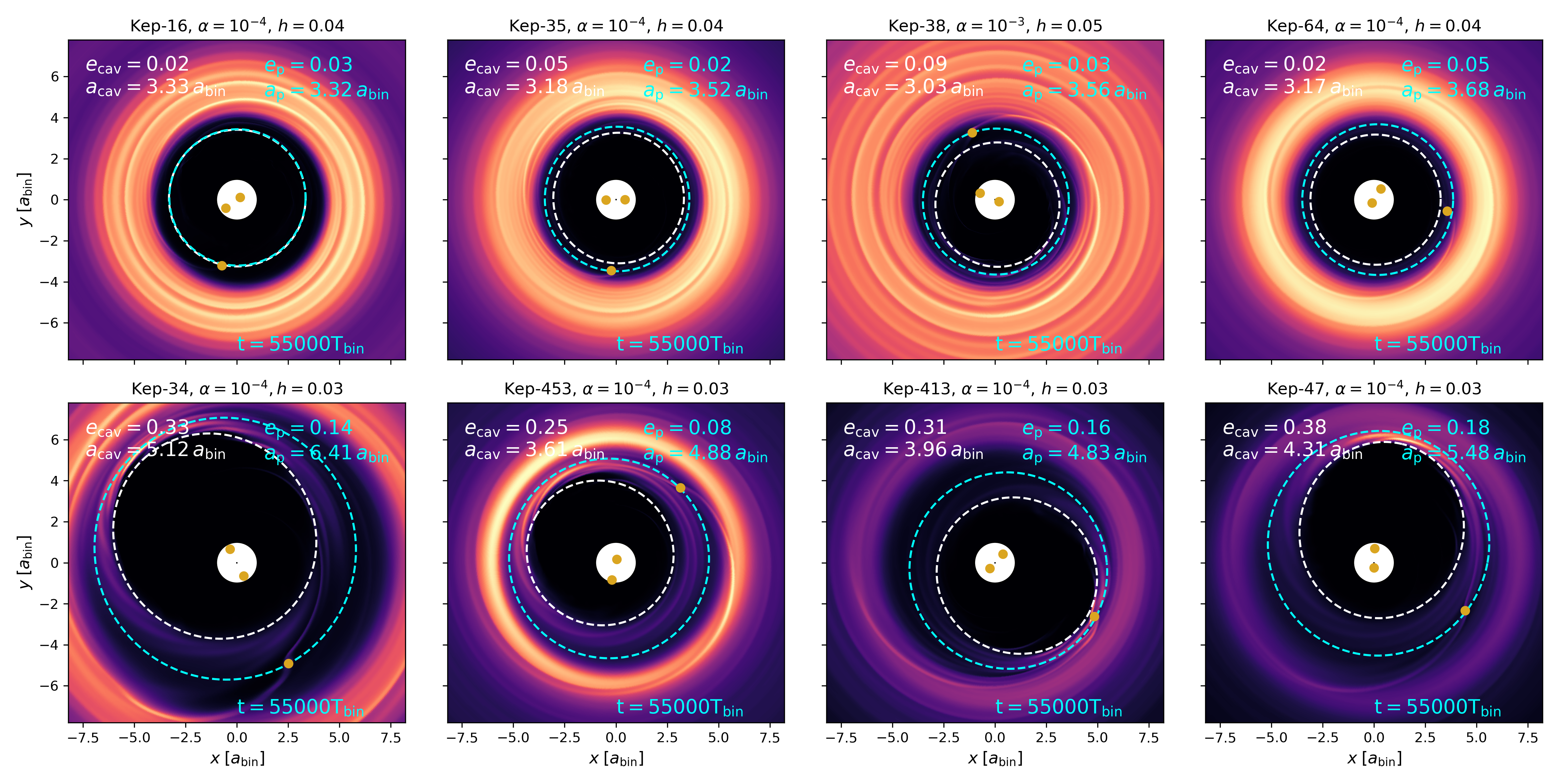}
    \caption{Discs and planets of 8 of the considered Kepler systems listed in Table \ref{tab:Kep-data} and shown in Fig.\,\ref{fig:discs} but now with embedded planets. The quoted times are measured after insertion of the planets. The chosen aspect ratio $h$ and viscosity $\alpha$ correspond to those models that most successfully could reproduce the observed planet orbits (see Table\,\ref{tab:Sim-result} below). The white dashed line marks the inner cavity edge and the light blue line the planet orbit. The top row shows planets that correspond to our success criteria of reaching the observed orbit by $\pm 0.1\mathrm{a_{obs}}$, while the lower row displays those systems that did not reach the observed orbit, due to their eccentric discs.}
    \label{fig:planets}
\end{figure*}

\begin{figure}[htb]
    \centering
    \includegraphics[width=0.5\textwidth]{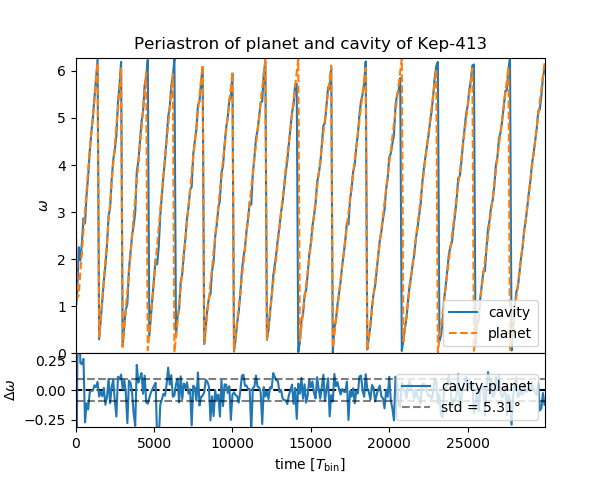}
    \caption{The argument of pericentre of the elliptic cavity and planet orbit for Kepler-413.
   The standard deviation of $\Delta \omega = | \omega_\mathrm{p} - \omega_\mathrm{cav} | $ is 5.31 degree.}
    \label{fig:peri}
\end{figure}

After the discs reached a quasi-equilibrium structure planets with the observed masses were embedded in the disc at a distance $\geq6 a_\mathrm{bin}$. 
Their subsequent migration histories are illustrated in Fig.\,\ref{fig:close_planets} 
for those six cases where the match between simulated and observed location was closest.
The systems with an intermediate value of $\mathrm{e_{bin}}$ shown good convergence to the observed orbits 
for an $\alpha$-value of $10^{-4}$ and aspect ratio of 0.04 to 0.03. We find here, that a low viscosity of $\alpha=10^{-4}$ is needed for the discs to allow the migration into the observed orbits. For Kepler-38b we performed an additional run with a higher aspect ratio $h=0.05$, because Kepler-38b is the most massive planet of the sample and, as shown in Fig.~\ref{fig:close_planets}, it reaches just the success-limit with $\alpha = 10^{-3},\, h=0.05$ and will cross the dashed line after $120\,000 T_\mathrm{bin}$. With this slightly higher pressure the density maximum of the disc is reduced leading to a reduce positive torque of the disc onto the planet, allowing it to park closer to the binary than with lower temperatures. Another important factor for this single planet is its high mass. As shown in Table \ref{tab:Sim-result} all other planet orbits stay further out for $\alpha = 10^{-3}, h=0.04$ than Kepler-38b.  Thereby, Kepler-16, -35, -38, -64, TOI-1338, and also the recently discovered Kepler-1661b \citep{2020Socia} can be simulated with a single parameter set of $\alpha \sim 10^{-4},\,h=0.04$ to produce a final orbit with an accuracy of $0.1\,\mathrm{a_{p,obs}}$ as shown in Fig.~\ref{fig:close_planets}.
As noticeable in Fig.\,\ref{fig:close_planets}, the initial migration in the outer disc is typically very fast because the planets are fully embedded in the disc and have not yet opened any gap. The systems Kepler-1661, -64 and TOI-1338b have a comparable migration rate as their planet to binary mass ratios are very similar. When they reach the high density ring at around $5\,\mathrm{a_{bin}}$ or $1.5 \,\mathrm{a_{obs}}$, the migration is slowed down. They are influenced by the high density inner disc region and the binary, and subsequently experience some variations in their semi-major axis and eccentricity evolutions. The first migration reversal occurs when they are getting close to a semi-major axis of about $4\,\mathrm{a_{bin}}$.


The two systems Kepler-16 and -38 show initially a very fast inward migration. As they have a higher planet-binary mass ratio than the other planets they would in the embedded type-I migration phase exhibit faster inward migration \citep{2012Kley}. Additionally, being in the Saturn mass range they are subject to rapid type-III migration \citep{2003ApJ...588..494M}. In the later evolutionary phase they begin to circularise the disc and can migrate inwards smoothly. This is especially evident in the unique model of Kepler-38 with high viscosity where the disc eccentricity drops from $e_\mathrm{cav}\approx0.37$ without planet to $e_\mathrm{cav}\approx0.09$ with an embedded planet.

Eventually all of them reach an orbit close to the inner cavity edge and stop their inward migration there because the migration torques vanish due to the changing disc structure. The cavity acts in the same way as a classic planet trap where a positive density gradient can stop a planet \citep{2006ApJ...642..478M}.
This can be seen in the top row of Fig.\,\ref{fig:planets} where we display the final surface density distribution for the
previously shown discs, now with embedded planets, after their migration has stopped.
Displayed are the orbital elements of the disc cavity and the planet, in the top left and right corners, respectively.
As discussed above, the ''successful'' evolutions have a final disc structure with small eccentricity, reduced by the presence of the planet.
The planets typically reach a final destination with nearly circular orbits at a distance of approximately 3.4 to 3.6 $\mathrm{a_{bin}}$,
top row in Fig.\,\ref{fig:planets}


As already shown in \cite{2018Thun}, binaries with very low and high eccentricities form highly eccentric inner cavities as shown in Fig. \ref{fig:discs}, bottom row. The planet has to overcome the density bulk at the cavity. This is more difficult for more eccentric cavities that have a higher peak density, and therefore for the high and low $\mathrm{e_{bin}}$ systems Kepler-34, -413, -435 and -47.
Figure \ref{fig:planets} shows the best models of these system with ''non-successful'' evolutions in the bottom row. While in the top row the planets reach a final position close to the observed one, in the bottom row the planets stall further out. Comparing the top to the lower systems the discs clearly show a different structure, namely that the planets around the high and low eccentric binaries are more embedded in the disc, remain at larger distances from the binary, and show a more eccentric orbit. In Fig. \ref{fig:planets} the blue and white ellipses in the bottom row show the same orientation which indicates that the disc-driven planet orbit becomes fully aligned with the eccentric cavity. This is shown for Kepler-413 in Fig. \ref{fig:peri}, the final pericentres of planet and disc orbits are aligned and the ellipses precess at the same rate, as also seen in \cite{2015Kley}, \cite{2019Kley} and \cite{2019Penzlin}. 
There are no clear signs of dynamical interaction (e.g. an oscillation of $\Delta \omega$) visible.
But the determination of the elements of the diskis quite noisy.
All the final orbits are aligned with the disc either embedded within the disc or located at the inner cavity edge.

\begin{figure}[t]
    \centering
    \includegraphics[width=0.5\textwidth]{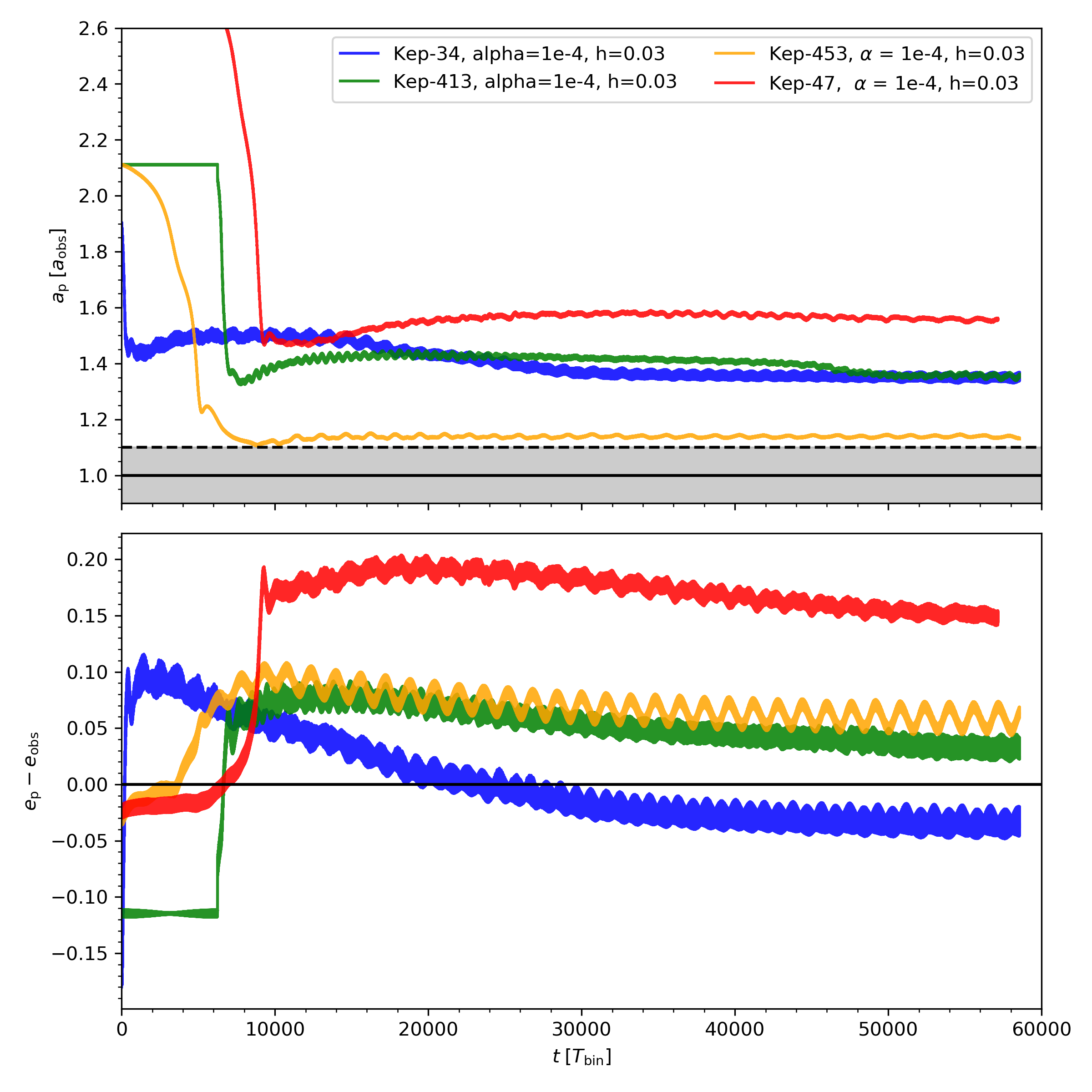}
    \caption{Migration history for all planets that do not get close enough to the observed orbit. Semi-major axis and eccentricity are scaled with the observed $\mathrm{a_{obs}}$ and $\mathrm{e_{obs}}$. The grey area marks the success criterion.}
    \label{fig:off_planets}
\end{figure}

The orbital evolution of the embedded planets in these ``unsuccessful'' systems, shown in Fig.~\ref{fig:off_planets}, starts with the same fast initial migration as in the previous cases, but then the inward migration is stalled by the disc too far out. Kepler-34, being the most eccentric binary with an eccentricity of 0.52, produces an eccentric disc with a large cavity and subsequently a planet on a large orbit. 

Kepler-453b has the furthest observed orbit of all CBPs (normalised to the binary separation) around low eccentric binaries, and it almost reaches the required orbit.
In this case the planet to binary mass ratio is very small such that the planet is not able to alter the disc structure considerably with the given disc parameters.
Kepler-453b has a four times smaller planet-binary mass ratio than Kepler-413b which is observed to be $0.75\, a_\mathrm{bin}$ further in. This hints qualitatively to the importance of the planet mass in reaching orbits closer to the binary as the dimensions of these systems are comparable. 
We will explore the effect of mass in detail in the next section, with models that only change the planet masses. However, in this simulations both are stopped by the disc rather than reaching an orbit inside the inner cavity. 
The planets' distant final location for systems with nearly circular binaries and low disc viscosity is here not necessarily caused by low planet masses because Kepler-413b has relatively a high mass with $0.21\,\mathrm{m_{jup}}$.

In order to test the hypothesis that the ability of the planet to open a gap is crucial to detach from the disc and reach the observed low eccentric orbits, we ran additional simulations for the low and high $e_\mathrm{bin}$ systems with planets of 2 and 10 times the observed best fit mass. Due to the transit observation method the uncertainty of the observed masses is high and the confidence interval reaches masses of around a factor of 2 of the best fit model. 
For the 10 times heavier model we will vastly overestimate the mass of the planet for the sake of the theoretical exploration and no longer work within the constraints of the observation. We will compare this to all other models in the following subsection.

\subsection{Model results and best-fit parameter set}

All simulation results are summarised in Tab.\,\ref{tab:Sim-result}, where we quote if a simulation has successfully
reproduced the observational parameter of a given planet. The systems are sorted by the eccentricity of the host binary. 
With the exception of the heaviest planet, Kepler-38b, the $\alpha=10^{-3}$ environment is not sufficient to explain the observations. 
Just by reducing the viscosity to $\alpha=10^{-4}$, we find that for 6 of 10 systems the final semi-major axis
deviates by no more than 10\% from the observed value, see the grey shaded region in 
Fig.~\ref{fig:close_planets}.

To understand now the reasons for the unsuccessful simulations we have to analyse the differences in the systems.
The first three binary systems have $\mathrm{e_{bin}}\leq 0.05$. The next six systems have $0.10 \leq \mathrm{e_{bin}} \leq 0.21$,
while the last system has $\mathrm{e_{bin}}= 0.52$. 
The result can be split into these three regimes.

For the nearly circular binaries it is difficult to match the observations. However, by reducing the disc scale height and the viscosity to $\alpha=10^{-4}$ and $h=0.03$ the planets are able to all migrate further inwards as they are more easily able to clear more of their orbit. In these systems the effect of a 2 fold planet mass increase is comparable to the impact of lowering the disc aspect ratio by 0.01.
With the 10 fold mass increase, a planet of $0.5\,\mathrm{m_{jup}}$ in the Kepler-47 system can migrate to the observed position. The other heavy planets with a 10 fold mass increase will migrate towards a less eccentric and closer-in position than the observed location and may potentially reach unstable orbits.

The second regime of intermediate eccentric binaries can easily be simulated with the set of $\alpha=10^{-4}$ and $h=0.04$. This is a good indication that the conditions in the disc should be comparable to this parameter set. In our previous study \citep{2019Kley} we have shown that in a viscously heated disc the inner disc scale height is unflared and close to $h=0.04$. 
Recent studies, observationally in the D-sharp survey \citep{2018Dsharp6} and theoretically \citep{2020Flock}, suggest that viscosities are $\alpha\leq10^{-3}$, and therefore our choice of a low viscosity is reasonable. The well matched final planet orbits can be taken as an additional indicator for low viscosities in protoplanetary discs.

The one observation with very high binary eccentricity Kepler-34 is hard to reproduce. With an increase in planet mass the correct semi-major axis of the planet could be reached, however the eccentricity is lowered to values below the observed one, and already too low in a system with a reduced scale height. 

\begin{table*}
\centering
\begin{tabular}{|c||c|c|c||c|c||c|}
\hline 
 Kepler- & $\alpha = 10^{-3}$,& $\alpha = 10^{-4},$& $\alpha = 10^{-4},$ &$\alpha = 10^{-4},$&$\alpha = 10^{-4},$& obs.\\ 
 System &  $h=0.04$ &  $h=0.04$ &  $h=0.03$ & $h=0.04,\,2\,m_p$ & $h=0.04,\,10\,m_p$ & $\mathrm{a_p},\, \mathrm{e_p}$ \\
\hline 
\hline
47 & $6.6\, \mathrm{a_{bin}},\, 0.23$ & $6.6\, \mathrm{a_{bin}},\, 0.23$ & $5.0\, \mathrm{a_{bin}},\, 0.18$ & $5.0\, \mathrm{a_{bin}},\, 0.18$ & \checkmark &$3.53\, \mathrm{a_{bin}},\, 0.03$\\ 
\hline 
413 & $6.0\, \mathrm{a_{bin}},\, 0.21$ & $5.7\, \mathrm{a_{bin}},\, 0.18$ & $4.8\, \mathrm{a_{bin}},\, 0.15$& $5\, \mathrm{a_{bin}},\, 0.15$&  $<3.1\, \mathrm{a_{bin}},\, 0.025$  &$3.55\, \mathrm{a_{bin}},\, 0.12$\\
\hline 
453 & eject & eject & $4.8\, \mathrm{a_{bin}},\, 0.12$ & $5.2\, \mathrm{a_{bin}},\, 0.15$ & $<3.1\, \mathrm{a_{bin}},\, 0.025$ & $4.26\, \mathrm{a_{bin}},\, 0.036$\\ 
\hline
38 &  $3.8\, \mathrm{a_{bin}},\, 0.02$ & \checkmark & & & & $3.16\, \mathrm{a_{bin}},\, 0.03$\\ 
\hline 
1661 & &\checkmark & & & & $3.39\, \mathrm{a_{bin}},\, 0.06$\\ 
\hline 
35 & $4.7\, \mathrm{a_{bin}},\, 0.1$ & \checkmark & & & & $3.43\, \mathrm{a_{bin}},\, 0.04$\\ 
\hline  
TOI-1338 & $4.2\, \mathrm{a_{bin}},\, 0.08$ & (\checkmark), $0.05$ & & & &$3.49\, \mathrm{a_{bin}},\, 0.09$\\
\hline 
16 & $4.1\, \mathrm{a_{bin}},\, 0.04$ & \checkmark & & & &$3.14\, \mathrm{a_{bin}},\, 0.006$ \\ 
\hline 
64 & $4.8\, \mathrm{a_{bin}},\, 0.08$ & \checkmark & \checkmark & & & $3.64\, \mathrm{a_{bin}},\, 0.05$\\ 
\hline 
34 & $>9\, \mathrm{a_{bin}},\, 0.15$ & $7.8\, \mathrm{a_{bin}},\, 0.18$ & $6.4\, \mathrm{a_{bin}},\, 0.17$ & $<4.5\, \mathrm{a_{bin}},\, <0.05$ &  $4.5\, \mathrm{a_{bin}},\, 0.025$  &$4.76\, \mathrm{a_{bin}},\, 0.18$\\ 
\hline
\end{tabular}
\caption{Check-table indicating if a simulation for a given viscous $\alpha$-value, aspect ratio $h$, and planet mass reached the observed orbit.
The systems are ordered from lowest $\mathrm{e_{bin}}$ to highest.
Every cell contains the semi-major axis and eccentricity of the final planet orbit of the stated simulation.
If not stated (columns 2 to 4), the planet mass corresponds to the observed mass in Tab.~\ref{tab:Kep-data}.
For boxes with a "\checkmark", the planet's final location lies within $0.1\,\mathrm{a_{bin}}$ of the observed position.
In columns 5 and 6 the planet mass in the simulations have been increased by a factor of 2 and 10, respectively. 
For comparison, the inferred parameters from the Kepler data are on the right. Blank fields were not simulated.}
\label{tab:Sim-result} 
\end{table*}

\section{Discussion \& Summary} \label{sec:dicussion}
We have performed 2D hydrodynamical simulations of circumbinary discs with embedded planets in order to explain the orbits of the
observed sample of circumbinary planets.
One simplification of the model is to use a locally isothermal disc. In previous work \citep{2019Kley}, we explored viscously heated, radiatively cooled discs. The results showed a plateauing aspect ratio at $0.03<h<0.05$ for the inner $15\, \mathrm{a_{bin}}$.
According to those results we chose a constant aspect ratio in this range, as this produced a reasonable and well-defined test disc.

Contrasting previous studies, where the planets typically were parked at too large distances from the binary,
our simulations show that we can explain the planetary orbits in systems which have a binary eccentricity between 0.05 to 0.3 by applying a simple locally isothermal model with a parameter set of a viscous $\alpha = 10^{-4}$, and a scale height of $H/r = 0.04$.
Variations in the binary mass ratio do not change the results in any significant way (as also found in \cite{2018Thun}). 
For these systems with intermediate binary eccentricity, planet to binary star mass ratios between $4.6\cdot 10^{-5}$ (Kepler-1661) and $30\cdot 10^{-5}$ (Kepler-38) are sufficient to carve out planetary gaps, move independently of the disc and settle to closer orbits.
Additionally, the small viscosity circularizes the disc even without a planet, which allows for tighter orbits as well.
Using this parameter set ($\alpha = 10^{-4}, h = 0.04$), we can reproduce the observed orbits of 6 of 10 Kepler circumbinary planets with hydrodynamical simulations,
and obtain a much closer agreement to the observations for the other four systems than previous studies. 

It is instructive to compare our findings to standard gap opening criteria derived for planets emdedded in discs around single stars.
The mass of a planet required to open a gap can be estimated by 
a thermal and viscous criterion. The thermal criterion requires the Hill radius to exceed the local pressure scale height
while the viscous criterion requires the gravitational torque to exceed the viscous one. In terms of the aspect ratio, the visocity $\alpha$
and the mass ratio $q_\mathrm{p} = \mathrm{m_p}/\mathrm{M_{bin}}$ they read \citep{1999ApJ...514..344B}
\beq
   q_\mathrm{p,th} > 3 h^3  
  \quad  \mbox{and} \quad 
   q_\mathrm{p,visc} > 40 \alpha h^2 \,.
\eeq
Up to factors of order unity these agree with other criteria \citep{2006Crida,2020Alex}.

For $\alpha = 10^{-4}$ and $h = 0.04$ one finds $q_\mathrm{p,th} \approx 2 \times 10^{-4}$ 
and $q_\mathrm{p,visc} \approx 6.4  \times 10^{-6}$.
While the viscous criterion is matched by all systems the thermal one is only fulfilled clearly for Kepler-38b and marginally
by Kepler-16b and Kepler-413b. This explains why for the Kepler-38 system a larger viscosity ($\alpha = 10^{-3}$) is still sufficient.

There have been a number of recent studies of the opening and structure of gaps due to planets around single stars ( \cite{2006Crida}; \cite{2013Duffell}; \cite{2014Fung}; \cite{2015Duffell}; \cite{2015Kanagawa,2016Kanagawa}). Given that circumbinary discs and planets become eccentric, planet-induced gap structures are likely to be different in this scenario. Undertaking a detailed study of this goes beyond the scope of this paper, but would make an interesting topic for future work.

The systems that could not be matched that well were those with very small and large binary eccentricities.
For both of these extreme cases, the inner cavity of the disc is most eccentric and the planets were not able to
to 'free' themselves from the dynamical forcing of the disc given
their masses are at most $0.2\, \mathrm{M_{jup}}$. This might be due to the fact that for these eccentric discs
higher planet masses are required to clear a gap. This is supported by simulations
with artificially increased planet masses where we observed orbital parameters in closer agreement with the observations.
The close, eccentric orbit around the highly eccentric binary (Kepler-34) is still difficult to explain by this simple approach of planet migration within a disc because the presence of the disc damped the planet eccentricity below the observed value. The study of possible further (tidal) interactions after the dissipation of the disc, however, exceeds the scope of this work.

Even though the thermal mass is nearly reached by the planet for the low $\mathrm{e_{bin}}$ Kepler-413 system, the final planet orbit 
is not in very good agreement with the observed one, which is probably due to the high disk eccentricity. In the other small $\mathrm{e_{bin}}$ system Kepler-453 the planet has a very low mass and a relatively large orbit. So for this planet it appears that the migration was halted by the disc earlier than for the comparable heavier planet in the low eccentricity binary like Kepler-413b.
Kepler-47 is the only observed multi-planet circumbinary system which may influence the migration.

The upper mass of close-in circumbinary planets in the observations is limited to $\sim 0.4\, \mathrm{M_{jup}}$. When we simulated the more massive planets, we observed that heavy planets migrate further in, so they might reach unstable orbits and suffer ejections, in agreement with \cite{2008Pierens}.

The binary parameters may also change over time. However, to compare to the existing systems we kept them fixed in the initial disc simulation. As the binaries are close, $10\,000 \, \mathrm{T_{bin}}$ is less than 2000 years. For the planet evolution simulations, the binary orbit
is also effected by the forces from the disc but changes in the binary parameters during this time frame are negligible, amounting to about $~1\%$ in $60\,000\, \mathrm{T_{bin}}$. The shrinkage of the binary orbit on longer timescales could be accounted for by starting the system at larger separations
than observed today, but that would not be possible by direct hydrodynamical simulations, and would lead to the same result eventually.

The reason why in some of the systems the final orbits remain too eccentric and too far out may also be the result of
missing physics from the models. In our models we neglected radiative transport vertically and within the disc's plane and these could
potentially reduce the disc's eccentricity. We have modelled the discs with 2D simulations but by moving to 3D one may obtain better agreement for the
systems because disc eccentricities/cavity sizes are different in 3D versus 2D. For example, in their 3D simulations, \citet{2018MNRAS.477.2547P} found 
that the planet in the difficult to match system Kepler-413 moved closer to the binary.

Possibly, multiple planets are present in some of the systems that are even slightly inclined so that we see only one of the planets. 
In \citet{2019Penzlin} we showed that a secondary planet does not change the orbit of the first one significantly, hence we did not include multi-planet simulations here.

It may also be possible to reduce the binary eccentricity through tidal interaction with the planet after disc has been dispersed. Thereby, the low eccentric binaries may have been more eccentric during planet formation phase and only lost eccentricity after the planets reached the stable orbits undisturbed by the outer disc \citep{2019tidal}.

After all, the fact that all planets in systems where the circumbinary discs have intermediate eccentricity could be matched well, shows
that the migration scenario of explaining short period circumbinary planets is very likely to be valid.
Only those systems with high eccentric discs and large cavities point towards missing physics, or 3D effects.

\begin{acknowledgements}
Anna Penzlin was funded by grant KL 650/26 of the German Research Foundation (DFG).
Richard Nelson acknowledges support from STFC through the Consolidated Grants ST/M001202/1 and ST/P000592/1 and from the Leverhulme Trust through grant RPG-2018-418.
The authors acknowledge support by the High Performance and
Cloud Computing Group at the Zentrum f\"ur Datenverarbeitung of the University
of T\"ubingen, the state of Baden-W\"urttemberg through bwHPC and the German
Research Foundation (DFG) through grant no INST 37/935-1 FUGG.
The plots in
this paper were prepared using the Python library matplotlib \citep{Hunter:2007}.
\end{acknowledgements}

\bibliography{cbp-orbit,wk}
\bibliographystyle{aa}
\end{document}